\documentstyle[seceq,supplement,epsf,amssymb,wrapft]{ptptex}

\def\ave#1{\langle #1\rangle}
\def\C{{\Bbb{C}}}
\def\Z{{\Bbb{Z}}}

\newcommand{\ul}[1]{\underline{#1}}

\newcommand{\Ord}[1]{{\cal O}\left(#1\right)}

\newcommand{\half}{{\textstyle{\frac{1}{2}}}}
\newcommand{\halpha}{{\textstyle{\frac{\alpha}{2}}}}
\newcommand{\ghalf}{{\textstyle{\frac{\gamma}{2}}}}
\newcommand{\hbeta}{{\textstyle{\frac{\beta}{2}}}}
\newcommand{\ihalf}{{\textstyle{\frac{i}{2}}}}
\newcommand{\shalf}{{\textstyle{\frac{1}{\sqrt{2}}}}}

\newcommand{\equar}{{\textstyle{\frac{\epsilon}{4}}}}

\newcommand{\sgn}{\,{\rm sgn}}
\newcommand{\tr}{\,{\rm tr}}
\newcommand{\ad}{\,{\rm ad}}
\newcommand{\ki}{{\rm\sc ki}}
\newcommand{\ih}{{\rm\sc i}}

\title{
Exact time-correlation functions of quantum Ising chain in 
a kicking transversal magnetic field
}
\subtitle{Spectral analysis of the adjoint propagator in Heisenberg picture}

\author{
Toma\v z {\sc Prosen}
}

\inst{
Max Planck Institut f\" ur Kernphysik, Postfach 103980, 62029 Heidelberg,
Germany, and
Physics Department, Faculty of Mathematics and Physics,
University of Ljubljana, Jadranska 19, 1111 Ljubljana, Slovenia
\footnote{Permanent address. E-mail address: prosen@fiz.uni-lj.si}
}

\recdate{
\today
}

\abst{
Spectral analysis of the {\em adjoint} propagator in a suitable
Hilbert space (and Lie algebra) of quantum observables in 
Heisenberg picture is discussed as an alternative approach to 
characterize infinite temperature dynamics of non-linear 
quantum many-body systems or quantum fields, and to provide a bridge 
between ergodic properties of such systems and the results of classical 
ergodic theory. We begin by reviewing some recent analytic and numerical 
results along this lines. In some cases the Heisenberg dynamics 
inside the subalgebra of the relevant quantum observables can be 
mapped explicitly into the (conceptually much simpler) Schr\" odinger 
dynamics of a single one-(or few)-dimensional quantum particle.
The main body of the paper is concerned with an application of the 
proposed method in order to work out explicitly the general spectral 
measures and the time correlation functions in {\em a quantum Ising 
spin $1/2$ chain in a periodically kicking transversal magnetic field}, 
including the results for the simpler autonomous case 
of a static magnetic field in the appropriate limit. The main result, 
being a consequence of a purely continuous non-trivial part of the 
spectrum, is that the general time-correlation functions decay to 
their saturation values as $t^{-3/2}$.
}

\begin{document}

\maketitle

\section{Introduction}
Quantitative and even qualitative understanding of dynamics and ergodic 
properties of interacting quantum many-body hamiltonian systems and fields 
(at finite (or infinite) temperature) is currently at its very early stage.
Situation is much better, if one considers {\em bounded} one or few particle 
systems, where ergodic theory\cite{cornfeld} (on a classical level) 
describes a variety of dynamical behaviours ranging from 
{\em complete integrability} being characterized by purely discrete 
spectrum of the appropriate evolution (Liouville) operator, thru 
ergodicity, mixing and chaos characterized by the continuous spectrum. 
Mixing, which is equivalent of saying that correlation functions of an 
arbitrary pair of observables ($A$,$B$) decay 
in time, $\lim_{t\rightarrow\infty}(\ave{A(t)B}-\ave{A}\ave{B})=0$,
is the necessary dynamical property needed to justify the relaxation
to equilibrium (micro-canonical) state and the laws of
statistical mechanics, such as the fluctuation-dissipation theorem 
and transport laws. We know that typical classical bound few body system 
is {\em intermediate}, neither completely integrable nor fully ergodic 
and mixing, and hence its evolution spectrum contains both, nontrivial
{\em point} spectral component connected to quasi-periodic dynamics
in regular parts of classical phase space described by the KAM
theorem, and the {\em continuous} spectral component connected to 
stochastic motion on chaotic components of phase space.
On a quantum level, bounded quantum systems of few particles have
always purely discrete spectrum and hence their time evolution is 
(asymptotically) quasi-periodic, so they can never be truly mixing, and their 
dynamical properties can only in (semi)classical limit $\hbar\rightarrow 0$
approach the ones  of their classical counterparts.
Another possibility of obtaining continuous evolution spectrum and truly 
mixing quantum behaviour is to consider {\em thermodynamic limit} where
the size (number of degrees of freedom) of the quantum many-body system
or field becomes infinite.
It has been suggested recently\cite{prelovsek1} that the finite temperature
dynamics of {\em integrable} quantum many-body lattices (solvable by 
{\em quantum inverse scattering} or {\em Bethe ansatz}) is pathological
(read: non-ergodic) from the point of view of statistical mechanics and 
transport phenomena, which may be explained by means of existence of an 
(infinite number of) exact conservation laws\cite{prelovsek2}.
On the other hand, several numerical studies\cite{JLP,P98ktv} of 
high temperature dynamics of strongly {\em non-integrable} quantum `many'-body 
systems of interacting particles suggested that their dynamics can
indeed approach mixing behaviour in thermodynamic limit.
However, it has been suggested in refs.\cite{P98ktv,P98inv},
based on numerical results in a family of kicked fermionic lattices,
that intermediate behaviour (of non-integrable but also non-ergodic and
non-mixing quantum dynamics) may exist as well in thermodynamic limit in a
finite range of systems' parameters.

In ref.\cite{P98inv} the Heisenberg dynamics of a certain Lie subalgebra
of quantum observables equipped with a Hilbert space structure 
of the above mentioned kicked fermion model on an infinite lattice
has been studied. 
It has been shown by numerical algebra in operator space that the regime of 
the so called intermediate dynamics, discovered before\cite{P98ktv} by direct 
Schr\" odinger time evolution on finite lattices, exactly corresponds with 
the existence of {\em few} (in contrast to {\em infinite} for integrable 
case) conservation laws, which are the eigenvectors (with eigenvalue 1) of 
the adjoint evolution propagator over the Hilbert space of quantum observables.

Later on\cite{P99map}, a similar program has been undertaken with more 
analytical approach: Dynamics of the adjoint propagator in Heiseberg picture
over the {\em two parametric} infinitely dimensional {\em dynamical} 
Lie-algebra of observables over 1d quantum spin $1/2$ chains, 
where the possibly time-dependent Hamiltonian can be any hermitean 
member of the algebra, has been formally mapped onto Schr\" odinger dynamics of a 
non-linear one-particle problem in 2d configuration space. Since the 
two spectral problems are shown to be equivalent, the (infinite-temperature) 
time auto-correlation functions of the spin-chains are identical to the 
quantum recurrence amplitudes of the associated one-body problem. 
Conceptually perhaps even more interesting is
the result that the continuum field limit of the 
spin-chains corresponds (or maps on) to the classical limit of the 
associated non-linear one-particle problem whose dynamics can go from 
integrable to truly mixing and chaotic.

In this paper we consider a related but simplified {\em one parametric} 
infinitely dimensional dynamical Lie algebra of spin chains which has been 
proposed in ref.\cite{P98ki} and used to construct infinite families of 
conservation laws for any member of the algebra being interpreted
as a Hamiltonian. Below we use and further develop these ideas 
in order to fully exploit the Heisenberg dynamics in the space of observables
(which is the framework that should correspond to the classical ergodic 
theory\cite{cornfeld} around Liouville equation), and exactly compute 
the time-correlation functions. We consider an interesting representative 
of this algebra, namely {\em the Ising chain periodically kicked with 
transversal magnetic field} $h$
\begin{equation}
H_\ki(t) = \sum_{j=-\infty}^{\infty}\left(
J\sigma^x_j \sigma^x_{j+1} + \delta_\tau(t) h \sigma^z_j\right),\quad
\delta_\tau(t):=\sum_{m=-\infty}^{\infty} \delta(t-m\tau).
\label{eq:ham}
\end{equation}
where $\sigma^p_j,p\in\{x,y,z\},j\in\Z$ are 
the standard spin $1/2$ (Pauli) operators at different sites $j$ 
satisfying the commutation relations 
$[\sigma^p_j,\sigma^r_k]=2i\delta_{jk}\sum_s\epsilon_{prs}\sigma^s_j$.
Note that in the continuous-time limit $\tau\rightarrow 0$ (\ref{eq:ham}) 
becomes an `ordinary' Ising chain in a static transversal field. 
In sec.\ref{sec:II} we review some known facts\cite{P98ki} about the algebra of 
Kicked Ising (KI) model. In sec.\ref{sec:III} we pose the spectral problem for
the adjoint propagator in Hilbert space over the algebra ob observables and
show its relation to correlation functions of infinite-temperature statistical
mechanics. Interestingly, the spectral problem for the adjoint propagator
can be formally interpreted in terms of a `quantum one-particle scattering
problem' on a 1d semi-infinite lattice.
First, in sec.\ref{sec:IV} the limiting case of 
ordinary Ising chain in a static field is solved explicitly, 
and then in sec.\ref{sec:V} general results are given for the 
spectral measures and explicit asymptotics for the 
time-autocorrelation functions, which are shown to decay to their saturation 
values as $t^{-3/2}$. However, we note a striking difference of the model 
in a static vs. periodically kicked field since the
limits $\tau\rightarrow 0$ and $t\rightarrow\infty$ do not commute as
is explicitly demonstrated in case of magnetization correlation function.

\section{Algebraic properties of Kicked Ising chain}
\label{sec:II}

We start with the so-called {\em dynamical} Lie (sub)algebra 
$\frak{S}$ of quantum observables over infinite spin 
chains\cite{P98ki} which is essentially generated by the two parts of 
the KI Hamiltonian, namely $\sum_j \sigma^x_j \sigma^x_{j+1}$ 
and $\sum_j \sigma^z_j$, and which is spanned by the two infinite
sequences of selfadjoint observables $U_n$ and $V_n$, $n\in \Z$,
namely
\begin{eqnarray}
U_n &=& \sum_{j=-\infty}^\infty
\left\{ \begin{array}{lll}
     &\sigma^x_j(\sigma^z_j)_{n-1}\sigma^x_{j+n}, & n \ge 1, \\
-\!\!\!\!&\sigma^z_j, & n = 0, \\
     &\sigma^y_j(\sigma^z_j)_{-n-1}\sigma^y_{j-n}, & n \le -1,
\end{array}\right. \label{eq:repr} \\
V_n &=& \sum_{j=-\infty}^\infty
\left\{ \begin{array}{lll}
     &\sigma^x_j(\sigma^z_j)_{n-1}\sigma^y_{j+n}, & n \ge 1, \\
     &1, & n = 0,\\
-\!\!\!\!&\sigma^y_j(\sigma^z_j)_{-n-1}\sigma^x_{j-n}, & n \le -1,
\end{array}\right. \nonumber
\end{eqnarray}
where $(\sigma^z_j)_k:=\prod_{l=1}^k \sigma^z_{j+l}$ for $k \ge 1$,
$(\sigma^z_j)_0:=1$, and satisfy
\begin{eqnarray}
\left[ U_{m},U_{n} \right] &=& 2i ( V_{m-n} - V_{n-m} ), \nonumber\\
\left[ V_{m},V_{n} \right] &=& 0, \label{eq:comrel}\\
\left[ U_{m},V_{n} \right] &=& 2i ( U_{m+n} - U_{m-n} ). \nonumber
\end{eqnarray}
One can turn the algebra ${\frak S}$ into the Hilbert space by defining
the following (canonical) scalar product of any pair $A,B\in {\frak S}$
\begin{equation}
(A|B) = \lim_{L\rightarrow\infty}\frac{1}{L 2^{L}}\tr_L (A^\dagger B),
\label{eq:metric}
\end{equation}
($\tr_L$ is a trace over the space of finite chains of length $L$) with respect to which
observables $\{U_n,V_n;n\in \Z\}$ form an ortho-normal (ON) basis.
Note that metric (\ref{eq:metric}) is {\em invariant}
with respect to the {\em adjoint map}, $(\ad A)B = [A,B]$,
namely $((\ad A^\dagger)B|C) = (B|(\ad A)C)$, meaning that a hermitean
observable $A^\dagger=A$ generates a hermitean adjoint operator $\ad A$
on the Hilbert space ${\frak S}$ of observables with respect to the metric 
(\ref{eq:metric}). The {\em canonical infinite temperature average} of 
some observable $A\in{\frak S}$ normalized by the 'volume' $L$ is just
$\ave{A} := (1|A)$.
In ref.\cite{P98ki} it has been shown that any Hamiltonian of the
general form $H=\sum_{m=-m_-}^{m_+} (h_m U_m + g_m V_m)$ is {\em integrable} 
since there exists analytic ${\frak S}$-valued function $T(\vec{\lambda})$ of 
$N=(m_+ + m_- + 1)$-tuple of complex variables
$\vec{\lambda}=(\lambda_1,\ldots,\lambda_N), |\lambda_n|<1$,
commuting with the Hamiltonian $[H,T(\vec{\lambda})]\equiv 0$ for any
$\vec{\lambda}$. From this procedure, two semi-infinite sequences of
independent and mutually commuting conservation laws have been determined, 
the (non-trivial) charges
$Q_k = \!\!\sum_{m=-m_-}^{m_+}\! \left[ h_m (U_{k+m}+U_{-k+m})
                         + g_m (V_{k+m}+V_{-k+m})\right]$,
and the (trivial) currents
$C_k = V_{k+1} + V_{-k-1}$, for $k=0,1,2\ldots$, where $Q_0=2H$.

Let us now turn to our KI Hamiltonian (\ref{eq:ham}), which we write as
$H_\ki(t) = J U_1 - h \delta_\tau(t) U_0$, or the {\em Floquet map} 
factorizing into the product of kick and `free' part
\begin{equation}
U_\ki = {\cal T}\exp\left(-i\int_{-0}^{\tau-0}\!\!dt H_\ki(t)\right) = 
\exp(-i\halpha U_1)\exp(i \hbeta U_0)
\end{equation}
where $\alpha := 2\tau J$, and $\beta := 2\tau h$.
The key object in this paper is the {\em adjoint propagator}
of observables in the Heisenberg picture (or the adjoint Floquet map)
\begin{equation}
U^{\ad}_\ki = \exp(-i\hbeta \ad U_0)\exp(i\halpha \ad U_1),\;\;
U^{\ad}_\ki A(m\tau) = U^\dagger_\ki A(m\tau) U_\ki = A((m\!+\!1)\tau).
\end{equation}
$U^{\ad}_\ki$ is a {\em unitary operator} over the space of observables
${\frak S}$, $(U^{\ad}_\ki A|U^{\ad}_\ki B)=(A|B)$. 
The algebra (\ref{eq:comrel}) yields a simple
evaluation of the exponentials of the adjoint generators 
$\exp(i\ghalf \ad U_m)A=e^{i\ghalf U_m}Ae^{-i\ghalf U_m}$,
namely
\begin{eqnarray}
\exp(i\ghalf \ad U_m) U_n &=&
c_\gamma^2 U_n + s_\gamma^2 U_{2m-n} + c_\gamma s_\gamma
(V_{n-m}\!-\!V_{m-n}),\nonumber\\
\exp(i\ghalf \ad U_m) V_n &=&
c_\gamma^2 V_n + s_\gamma^2 V_{-n} - c_\gamma s_\gamma 
(U_{m+n}\!-\!U_{m-n}).
\end{eqnarray}
where a shorthand notation $c_\gamma:=\cos\gamma,s_\gamma:=\sin\gamma$ is 
introduced. It turns out\cite{P98ki} that a similar algebraic construction of 
$T(\vec{\lambda})$ and a complete set of conservation laws as in the
Lie algebra is possible also in the corresponding {\em Lie group}. One finds
\begin{equation}
Q_k = s_\alpha c_\beta (U_{k+1} + U_{-k+1}) 
- c_\alpha s_\beta (U_k + U_{-k})
+ \half s_\alpha s_\beta (V_{k+1} + V_{-k+1} - V_{k-1} - V_{-k-1}),
\label{eq:conslaws}
\end{equation}
and the trivial currents, $C_k,\;k\ge 0$, which
are the {\em eigenstates} of the adjoint Floquet map with eigenvalue 
$1$, $U^{\ad}_\ki Q_k = Q_k,\;U^{\ad}_\ki C_k = C_k$, and generalize the
known conservation laws for the static field\cite{grady}.
$C_k$ commute with any other element of the algebra ${\frak S}$
and hence span the {\em maximal ideal} ${\frak I}$ of the algebra ${\frak S}$. 
In the following, we will subtract this trivial orthogonal subspace 
${\frak I}$ and consider the Heisenberg dynamics on the {\em derived} 
algebra ${\frak S}' = [{\frak S},{\frak S}] ={\frak S}-{\frak I}$. 

\section{Spectral measures and time-correlations of kicked
Ising chain}
\label{sec:III}

Let us now consider the spectral problem for the unitary adjoint 
propagator $U^{\ad}_\ki$. The problem is to find, for any element 
$A$ of the Hilbert space ${\frak S}'$, the corresponding
spectral measure $d\mu_A(\vartheta)$ on a unit circle 
$\vartheta\in [-\pi,\pi)$, such that for a suitable test function 
$f(z)$ one has the identity (see e.g. part III and appendix 2 of 
ref.\cite{cornfeld})
\begin{equation}
(A|f(U^{\ad}_\ki)A) = \int d\mu_A(\vartheta) f(e^{i\vartheta}).
\label{eq:decomp}
\end{equation}
The measure $d\mu_A(\vartheta)$ is composed of
(a series of) delta functions for the point spectral component and
of a continuous distribution function for the absolutely continuous part of 
the spectrum, and even of a multifractal distribution in case
of a singular continuous spectral component. It has been shown that
all the three spectral parts can coexist, for example for the Schr\" odinger 
problem of the quantized kicked Harper model \cite{kharper}.
However, we will show below that the Heisenberg propagator of KI model 
has only the trivial point spectrum $\vartheta=0$ corresponding
to the conservation laws (\ref{eq:conslaws}) and the absolutely continuous
spectrum with the continuous spectral measure 
$\mu'_A(\vartheta) = d\mu^{\rm a.c.}_A(\vartheta)/d\vartheta$.
Then, taking $f(z) = z^m$, we write the infinite temperature
autocorrelation function as a Fourier transformation of the 
spectral measure
\begin{equation}
\ave{A(m\tau)A^\dagger} = (A|[U^{\ad}_\ki]^m A) = 
D_A + \int_{-\pi}^\pi d\vartheta \mu'_A(\vartheta) e^{im\vartheta}.
\label{eq:tcor}
\end{equation}
Note that $\ave{A}=0$ for any $A\in \frak{S}'$. 
$D_A$ is the {\em time-averaged autocorrelation function} 
$D_A=\lim_{M\rightarrow\infty}(1/2M)\sum_{m=-M}^M \ave{A(m\tau)A^\dagger}$, 
or the {\em weight of the point spectral component} and can be computed
from the `sum-rule' (putting $m:=0$ in eq.(\ref{eq:tcor}))
$D_A =  (A|A) - \int_{-\pi}^\pi d\vartheta \mu'_A(\vartheta)$.
Nevertheless, $D_A$ can also be computed from the
full set of eigenstates --- {\em orthogonalized} conserved charges $Q'_k$
(obtained by applying the Gram-Schmidt orthogonalization onto the sequence 
$Q_k$ (\ref{eq:conslaws})),
$(Q'_k|Q'_l)=\delta_{kl}$, namely
\begin{equation}
D_A = \sum_{k} |(Q'_k|A)|^2,
\end{equation}
which is the essence of theorems on bounds for susceptibilities\cite{mazur}.

Let us organize the ON-basis of ${\frak S}'$ in the following way:
let $E_0:=U_0$ and $\vec{E}_n$ be the triple 
$\left(U_n,U_{-n},(V_n-V_{-n})/\sqrt{2}\right)$ for $n\ge 1$. 
Then general observable $A\in {\frak S}'$ can be expanded as
$A = a_0 E_0 + \sum_{n=1}^\infty \vec{a}_n\cdot\vec{E}_n =: \ul{a}\cdot\ul{E}$ 
with one scalar and a sequence of vector coefficients denoted by
$\ul{a} = (a_0,\vec{a}_1,\vec{a}_2,\ldots)$.
In the basis $\ul{E}$ the matrix of adjoint map $U^{\ad}_\ki$ can
be written as a banded ($3\times 3$ block-pentadiagonal) matrix 
$\ul{U}^{\ad}_\ki$ with a periodic structure except for small indices of 
rows/columns. As a consequence of this structure, 
the spectral problem for the generalized eigenfunctions
$\ul{\psi}(\vartheta)$, 
\begin{equation}
\ul{U}^{\ad}_\ki \ul{\psi}(\vartheta) = e^{i\vartheta}\ul{\psi}(\vartheta),
\label{eq:spcproblem}
\end{equation}
can be written uniquely as a `quantum mechanical one-particle 
scattering problem' on a semi-infinite 1d lattice, 
with the asymptotic part, for $n>2$,
\begin{equation}
A_\beta\left(B_\alpha \vec{\psi}_{n+2} 
       - C_\alpha \vec{\psi}_{n+1}
       + F_\alpha \vec{\psi}_n
       + C^T_\alpha \vec{\psi}_{n-1}
       + B^T_\alpha \vec{\psi}_{n-2}\right) = e^{i\vartheta} 
\vec{\psi}_n
\label{eq:asymp}
\end{equation}
where the $3\times 3$ matrices $A_\beta,B_\alpha,C_\alpha,F_\alpha$ read
\begin{eqnarray}
A_\beta &=& \pmatrix{c_\beta^2 & s_\beta^2 & -\sqrt{2}c_\beta s_\beta\cr
                   s_\beta^2 & c_\beta^2 & \sqrt{2}c_\beta s_\beta\cr
    \sqrt{2}c_\beta s_\beta & -\sqrt{2}c_\beta s_\beta & c_{2\beta}},\label{eq:matrices}\\
B_\alpha &=& s^2_\alpha \pmatrix{ 0 & 0 & 0 \cr
                                1 & 0 & 0 \cr
                                0 & 0 & 0 }, \quad
C_\alpha = \sqrt{2}c_\alpha s_\alpha \pmatrix{ 0 & 0 & 0 \cr
                                                 0 & 0 & 1 \cr
                                                 1 & 0 & 0 }, \quad
F_\alpha = \pmatrix{ c^2_\alpha & 0 & 0 \cr
                       0 & c^2_\alpha & 0 \cr
                       0 & 0 & c_{2\alpha} }, \nonumber
\end{eqnarray}
and with the `scattering potential at the origin' given by the equations
\begin{equation}
\pmatrix{ 
c^2_\alpha-e^{i\vartheta} & 0 & 0 & -\sqrt{2}c_\alpha s_\alpha & 
s^2_\alpha & 0\ldots \cr 
0 & c_{2\alpha} - c^2_\beta e^{i\vartheta} & -s^2_\beta e^{i\vartheta} &
-\sqrt{2}c_\beta s_\beta e^{i\vartheta} & 0 & 0 \ldots \cr
\sqrt{2}c_\alpha s_\alpha & \sqrt{2}c_\beta s_\beta e^{i\vartheta}
& -\sqrt{2}c_\beta s_\beta e^{i\vartheta} & c_{2\alpha} - c_{2\beta}e^{i\vartheta} &
-\sqrt{2}c_\alpha s_\alpha & 0\ldots} \ul{\psi} = 0.
\label{eq:scbc}
\end{equation}
(\ref{eq:scbc}) are only equations 1,2,4 out of the first 7 rows 
($n\le 2$) of the matrix equation 
$(\exp(i\halpha\ad U_1) - \exp(i\hbeta\ad U_0)e^{i\vartheta})\ul{\psi}=0$ 
which is equivalent to (\ref{eq:spcproblem}). The rows 3,6,7 are included 
already in the asymptotic part (\ref{eq:asymp}) 
(for $n=1,2$ since matrices $B_\alpha$ and $C_\alpha$ 
have many zero entries), while row 5 is equivalent to row 4,
and therefore they do not scatter the asymptotic solutions which are studied 
right below.

Now we solve the asymptotic problem (\ref{eq:asymp}) with the standard ansatz
leading to 
\begin{equation}
\vec{\psi}_n = \lambda^n \vec{v}(\lambda),\quad 
G(\lambda)\vec{v}(\lambda) = e^{i\vartheta}\vec{v}(\lambda)
\label{eq:asansatz}
\end{equation}
with the {\em transfer matrix} 
\begin{equation}
G(\lambda) = A_\beta 
(\lambda^2 B_\alpha - \lambda C_\alpha + F_\alpha 
+ \lambda^{-1} C_\alpha^T +\lambda^{-2} B_\alpha^T).
\end{equation}
The secular equation $\det(G(\lambda)-e^{i\vartheta}1_3)=0$ 
reduces to a simple fourth order {\em real} polynomial equation in 
$\lambda$, since 
$\det(G(\lambda)-e^{i\vartheta}1_3) = 
-ie^{3i\vartheta/2}\sin(\vartheta/2)\{a + b (\lambda + \lambda^{-1})+$ 
$c (\lambda^2 + \lambda^{-2})\}$, 
where $a = 1-c_{2\alpha}-c_{2\beta}-3c_{2\alpha}c_{2\beta}+4\cos{\vartheta}$,
$b=-2s_{2\alpha}s_{2\beta}$, 
$c=-2s^2_{\alpha}s^2_{\beta}$, with a simple solution which we write in terms of
two quadratic equations
\begin{equation}
\lambda + \lambda^{-1} = 
-2\frac{c_\alpha c_\beta \pm \cos(\vartheta/2)}{s_\alpha s_\beta}.
\label{eq:sqeq}
\end{equation}
Note, however that secular determinant is identically zero if the
spectral parameter vanishes $\vartheta=0$. Then, any function of the form
(\ref{eq:asansatz}) with $|\lambda|<1$ is a candidate for an eigenvector
of $U^{\ad}_\ki$ (or `bound state') and it has been shown\cite{P98ki} that linear
combinations of three of them generally solve the boundary equations (\ref{eq:scbc}), 
and from these a complete set of (local) conservation laws (\ref{eq:conslaws}) 
has been derived. 

In the following we thus exclude the `trivial' eigenvalue, and fix the
spectral parameter $\vartheta\ne 0$. $\lambda$ being a solution of 
(\ref{eq:sqeq}) means that $1/\lambda$ is also a solution. We 
interpret this as conservation of the magnitude of momentum, 
calling $\lambda$ a {\em momentum parameter}. The corresponding 
eigenfunction $\vec{v}(\lambda)$, 
will be {\em normalized} for convenience,
\footnote{$\vec{v}(\lambda)$ has been calculated explicitly using {\em Mathematica}, 
as all the other heavy algebraic calculations reported in this paper, 
and in general case is too lengthy to write down. We write it
explicitly later in various asymptotic regimes 
(\ref{eq:scdat1},\ref{eq:scdat2},\ref{eq:scdat3}).}
namely $\vec{v}(\lambda)^*\cdot\vec{v}(\lambda) = |\vec{v}(\lambda)|^2 = 1$.
Note that the eigenfunction $\vec{v}(\lambda)$ satisfies an interesting relation, 
namely $\vec{v}(\lambda)\cdot\vec{v}(\lambda^{-1})=0$ which can be proved from 
the following general property of the transfer matrix, namely 
$G^T(\lambda^{-1}) G(\lambda) = 1_3$ for any $\lambda\in\C$.

Further, it was shown (by means of computer algebra) 
that the `the scattering boundary condition' (\ref{eq:scbc}) 
can be solved for any $\lambda$ satisfying (\ref{eq:sqeq}) with 
the scattering ansatz of an incoming wave of an amplitude $1$ and outgoing wave
of an amplitude $S(\lambda)$,
\begin{equation}
\vec{\psi}_n = \vec{v}(\lambda^{-1})\lambda^{-n} + 
S(\lambda) \vec{v}(\lambda)\lambda^n,
\label{eq:psi}
\end{equation}
where the scattering amplitude reads
\begin{equation}
S(\lambda) = -\lambda^{-2} 
\frac{\vec{w}\cdot\vec{v}(\lambda^{-1})}{\vec{w}\cdot\vec{v}(\lambda)}, 
\quad{\rm with}\quad
\vec{w} = (c_\beta^2-e^{-i\vartheta},s_\beta^2,\sqrt{2}c_\beta s_\beta),
\quad
\label{eq:S}
\end{equation}
\begin{wrapfigure}{r}{6.5cm}
\epsfxsize = 6.8cm
\vspace{-4mm}
\centerline{\epsfbox{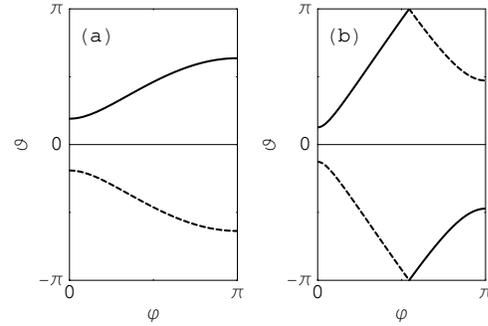}}
\vspace{-4mm}
\caption{Spectral bands ({\em upper} - solid, and {\em lower} - dashed curves) for
two cases: (a) $\alpha=0.35$,$\beta=0.65$, and (b) $\alpha=1.3$,$\beta=1.1$
(overlapping bands).}
\label{fig:band}
\end{wrapfigure}
and one finds the `unitarity condition for the S-matrix', 
$|S(\lambda)|=1$, for either real or complex $\lambda$. 
Apart from the scattering amplitude, 
boundary equations (\ref{eq:scbc}) also determine the scalar 
\begin{equation}
\psi_0 = v_2(\lambda^{-1}) + S(\lambda) v_2(\lambda).
\label{eq:psi0}
\end{equation}
From the above we learn that attenuating wave (\ref{eq:asansatz}) with
real momentum parameter $\lambda$, with $|\lambda^{-1}| < 1$, cannot generate an 
eigenstate of $U^{\ad}_\ki$ since it is {\em always}
accompanied with exponentially growing wave, $|\lambda|> 1$, due to the fact that 
$S(\lambda)$ cannot vanish. Therefore, {\em there is no point spectrum for} 
$\vartheta\neq 0$.
However, spectral parameters $\vartheta$ which admit complex unimodular
solutions of (\ref{eq:sqeq}) are in the absolutely continuous spectrum of 
the adjoint Floquet map $U^{\ad}_\ki$ with 
$\ul{\psi}$ (\ref{eq:psi}-\ref{eq:psi0}) being the generalized eigenvectors
in basis $\ul{E}$.
Writing $\lambda=e^{i\varphi}$ in terms of {\em quasi-momentum} 
$\varphi\in[0,\pi]$, putting it to (\ref{eq:sqeq}) and solving it for 
$\vartheta$, we obtain explicit forms of two continuous `bands' 
$\pm\vartheta(\varphi)$ with
\begin{equation}
\vartheta(\varphi) = 
2\arccos(\cos\alpha \cos\beta + \sin\alpha\sin\beta\cos\varphi) \pmod{2\pi},
\label{eq:bands}
\end{equation}
which go from $\pm2|\alpha-\beta|$ for $\varphi=0$ to 
$\pm2|\alpha+\beta|$ for $\varphi=\pi$, assuming without loss of generality 
that $|\alpha|,|\beta|<\pi/2$ because the full problem 
(\ref{eq:asymp}-\ref{eq:scbc}) is periodic in $\alpha,\beta$ with period $\pi$. 
Since the spectral parameter is on a unit circle
the two bands overlap for sufficiently large kick parameters, namely if 
$\max\{|\alpha+\beta|,|\alpha-\beta|\} > \half\pi$. In
this case, one has two different quasi-momenta $\varphi$ for a given fixed $\vartheta$
since eq.(\ref{eq:sqeq}) has two pairs of complex unimodular solutions.
See fig.\ref{fig:band}. The spectral measure of some observable $A\in {\frak S}'$, 
namely $\mu'_A(\vartheta)$, is non-vanishing only {\em inside the bands}, and it may be 
rewritten in terms of {\em quasi-momentum densities} on the bands $\rho^\pm_A(\varphi)$, 
\begin{equation}
d\mu_A(\pm\vartheta(\varphi)) = \rho^\pm(\varphi)d\varphi,\qquad
\mu'_A(\pm\vartheta(\varphi)) = 
|d\varphi/d\vartheta|\rho^\pm_A(\varphi),
\label{eq:rho2mu}
\end{equation} 
where in the last equation 
the two terms have to be added if the bands overlap, and we need the 
`density of states' 
$|d\varphi/d\vartheta|=|\sin(\half\vartheta)/(2s_\alpha s_\beta \sin(\varphi))|$.

\begin{figure}
\epsfxsize = 14.5cm
\vspace{-1truein}
\centerline{\epsfbox{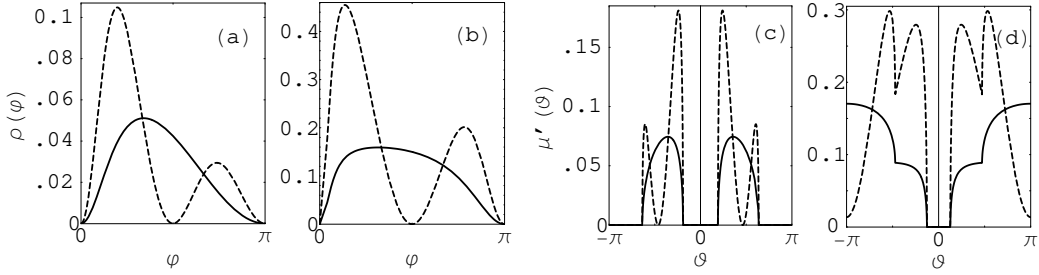}}
\vspace{-1truein}
\caption{Quasi-momentum densities (a,b) and spectral measures (c,d) for
the magnetization (solid curves) and XX-energy (dashed). (a,c) and (b,d)
are for cases (a) and (b) of fig.\ref{fig:band}, respectively.}
\label{fig:meas}
\end{figure}

Distributions $\rho^\pm_A(\varphi)$ can be calculated by means of a simple-minded
truncation of the Hilbert space ${\frak S}$ at $n=N$ and 
using counting-of-states-in-a-box technique yielding
\begin{equation}
\rho^\pm_A(\varphi)=\lim_{N\rightarrow\infty}\frac{N}{\pi}
\frac{|a^*_0\psi_0 + \sum_{n=1}^N \vec{a}^*_n\cdot\vec{\psi_n}|^2}
{|\psi_0|^2 + \sum_{n=1}^N \vec{\psi}^*_n\cdot\vec{\psi_n}}.
\end{equation} 
Using expressions (\ref{eq:psi}-\ref{eq:psi0}) we obtain,
writing $\vec{v}^\varphi:=\vec{v}(e^{i\varphi})$, $S^\varphi:=S(e^{i\varphi})$
\begin{equation}
\rho^\pm_A(\varphi)=
\frac{1}{2\pi}\Bigl|a^*_0 (v^{-\varphi}_2 + S^{\varphi}v^\varphi_2)
+ \sum_{n=1}^\infty \vec{a}_n^*\cdot
(\vec{v}^{-\varphi}_2 e^{-in\varphi} + S^{\varphi}\vec{v}^\varphi_2 e^{in\varphi})\Bigr|^2
\;\; {\rm with}\;\; \vartheta:=\pm\vartheta(\varphi).
\label{eq:rho}
\end{equation}
We notice that $\rho^+_A(\varphi) \equiv \rho^-_A(\varphi)$ iff $A=A^\dagger$.
We can finally transform the quasi-momentum densities back to the spectral measures,
or we write the spectral decomposition (\ref{eq:decomp}) directly in terms
of quasi-momentum integrals
\begin{equation}
(A|f(U^{\ad}_\ki) A) 
= D_A f(1) + 
\int_0^\pi d\varphi\left\{\rho^+_A(\varphi)f(e^{i\vartheta(\varphi)})
                         +\rho^-_A(\varphi)f(e^{-i\vartheta(\varphi)})\right\}.
\label{eq:decomprho}
\end{equation}
For example, we compute the total spectral weight of a point spectrum
\begin{equation}
D_A = (A|A) - \int_0^\pi d\varphi\left\{\rho^+_A(\varphi)+\rho^-_A(\varphi)\right\}.
\label{eq:DA}
\end{equation}
\begin{wrapfigure}{r}{6.5cm}
\epsfxsize = 6.8cm
\vspace{-2mm}
\centerline{\epsfbox{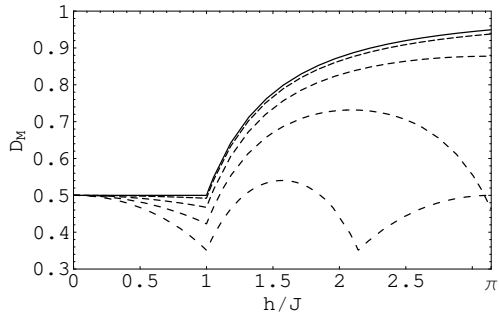}}
\vspace{-2mm}
\caption{
Dynamical susceptibility $D_M$ vs. relative magnetic field $h/J$.
Different curves, from solid to more and more open dashed curve, refer to different
values of $\tau=0$,$0.25$,$0.5$,$0.75$,$1$ (in this order).
}
\label{fig:stif}
\end{wrapfigure}
Formulae (\ref{eq:psi}-\ref{eq:psi0},\ref{eq:rho2mu},\ref{eq:rho},\ref{eq:decomprho}) are 
very useful exact results which we can use with some elementary numerics in order to 
compute the spectral measures $\mu'_A(\vartheta)$ and time correlation functions 
$(A|A(m\tau))$. In fig.\ref{fig:stif} we show results of calculation of the 
{\em dynamical susceptibility} $D_M$, namely the
time-averaged autocorrelation function of the magnetization 
$M=\sum_j \sigma^z_j = -U_0=\ul{m}\cdot\ul{E}$, $\ul{m}=(1,0,0\ldots)$
as the function of the relative field strength $\beta/\alpha=h/J$ 
(for several values of the kicking period $\tau$). Note an interesting singularity at 
$h/J=1$, or more generally at 
$\alpha=\beta\pmod{\pi}$, which will be commented on later in sec.\ref{sec:V}. 
In fig.\ref{fig:meas} we show quasi-momentum densities $\rho^+_A(\varphi)$ and
spectral measures $\mu'_A(\vartheta)$ for two observables, namely the 
magnetization $M=-U_0$ and the XX-chain Hamiltonian, 
$X=\sum_j(\sigma^x_j\sigma^x_{j+1} + \sigma^y_j \sigma^y_{j+1})=U_1+U_{-1}$,
$\ul{x}=(0,1,1,0,0\ldots)$, and for two
different sets of parameters $\alpha,\beta$ (same as in fig.\ref{fig:band}, 
with and without band overlap).
Moreover, one can obtain really explicit analytic results in two cases: 
(i) in the continuous-time limit $\tau\rightarrow 0$ of a static transversal 
magnetic field, and (ii) for asymptotically large times $t=m\tau\gg 1$.

\section{The limit of Ising chain in a static transversal field}
\label{sec:IV}

In the continuous time limit $\tau\rightarrow 0$, $t=m\tau$, 
we set the energy scale by putting
$J:=1$ ($\alpha=2\tau$, $\beta=2h\tau$), so we are left with a single parameter $h$.
$U^{\rm ad}_\ki=1 + i\tau\ad H_\ih + \Ord{\tau^2}$ is now an infinitesimal adjoint
propagator generated with the Hamiltonian $H_\ih = U_1 - h U_0$. One can again
formulate the spectral problem for the hermitean operator $\ad H_\ih$ in ${\frak S}'$ as 
(now somewhat simpler) one-particle 1d scattering problem. Or, one merely
expands general results (\ref{eq:psi}-\ref{eq:rho}) of the previous section for small 
$\tau$ and takes the limit $\tau\rightarrow 0$, writing the new spectral parameter of 
the hermitean operator $\ad H_\ih$ as $\epsilon=\vartheta/\tau$. Thus one finds the bands
$\pm\epsilon(\varphi)$,
\begin{equation}
\epsilon(\varphi) = 4\sqrt{1 + h^2 - 2h\cos\varphi}
\end{equation}
which extend from $\pm 4|h-1|$ to $\pm 4|h+1|$,
the `scattering data'
\begin{equation}
\vec{v}^\varphi = 
\left(2i(h - e^{i\varphi})/\epsilon(\varphi),\;2i(e^{-i\varphi}-h)/\epsilon(\varphi),\;
1/\sqrt{2}\right),\quad
S^\varphi\equiv -1,
\label{eq:scdat1}
\end{equation}
and the density of states $d\varphi/d\epsilon=\epsilon/(16h\sin\varphi)$.
Plugging all that into (\ref{eq:rho},\ref{eq:rho2mu},\ref{eq:decomprho}) we get
explicit formulae for any observable $A$.
Results are particularly simple (and perhaps physically interesting) 
for the magnetization $M$ where we find
\begin{equation}
\rho^\pm_M(\varphi) = \frac{\sin^2(\varphi)}{2\pi(1 + h^2 - 2h\cos\varphi)},
\;\;\mu'_M(\epsilon) = 
\frac{1}{4\pi h^2}\sqrt{((\equar)^2-(h\!-\!1)^2)((h\!+\!1)^2-(\equar)^2)}.
\label{eq:rhoM}
\end{equation}
It is very important to note the square-root singularities of the spectral measure 
$\mu'_M(4|h-1|+\varepsilon) \propto \varepsilon^{1/2}$,
$\mu'_M(4|h+1|-\varepsilon) \propto \varepsilon^{1/2}$, as $\varepsilon\searrow 0$,
since these will dominate the long-time behaviour 
of the infinite-temperature time-correlation function
\begin{equation}
\ave{M(t)M} = D_M + \frac{1}{4\pi h^2}\int_{4|h-1|}^{4|h+1|}\!\!\!d\epsilon 
\cos(\epsilon t)\sqrt{((\equar)^2-(h\!-\!1)^2)((h\!+\!1)^2-(\equar)^2)}.
\label{eq:MMt}
\end{equation}
We use an elementary asymptotics (which can be proved by complex rotation)
\begin{equation}
\int_a^{\pm\infty}\!\!dx \sqrt{|x-a|} f(x) e^{ixt} = \pm\frac{\sqrt{\pi}}{2}f(a)
\exp(iat\pm{\textstyle\frac{3\pi}{4}}i \sgn\,t)|t|^{-\frac{3}{2}} + 
\Ord{t^{-\frac{5}{2}}},
\label{eq:asymptotics}
\end{equation}
where $f(x)$ is some analytic function, applied to both ends of the spectrum in order to 
estimate the integral (\ref{eq:MMt}) giving (again, up to $\Ord{t^{-5/2}}$)
\begin{equation}
\ave{M(t)M} \approx D_M + 
\frac{
|h\!+\!1|^{\frac{1}{2}}\sin(4|h\!+\!1|t\!-\!{\textstyle\frac{\pi}{4}}) - 
|h\!-\!1|^{\frac{1}{2}}\sin(4|h\!-\!1|t\!+\!{\textstyle\frac{\pi}{4}})}
{16\sqrt{\pi}|ht|^{3/2}}.
\label{eq:magn}
\end{equation}
Further, the infinite-temperature dynamical susceptibility $D_M$ is
computed explicitly by integrating (\ref{eq:DA}) the distribution (\ref{eq:rhoM})
\begin{equation}
D_M(h) = 1 - \half (\max\{1,|h|\})^{-2},
\end{equation}
having the singularity at $h=h_c:=1$.
Note that the time average $D_M$ of $\ave{M(t)M}$ actually agrees with the
$\tau\rightarrow 0$ limit of the general case (see fig.\ref{fig:stif}, solid curve),
although the limits $\tau\rightarrow 0$ and $t=m\tau\rightarrow \infty$ do not 
generally commute as we show explicitly later for the correlation function 
$\ave{M(t)M}$ itself.

\section{Asymptotic results in general case}
\label{sec:V}

\begin{figure}
\epsfxsize = 14.5cm
\vspace{-2.5mm}
\centerline{\epsfbox{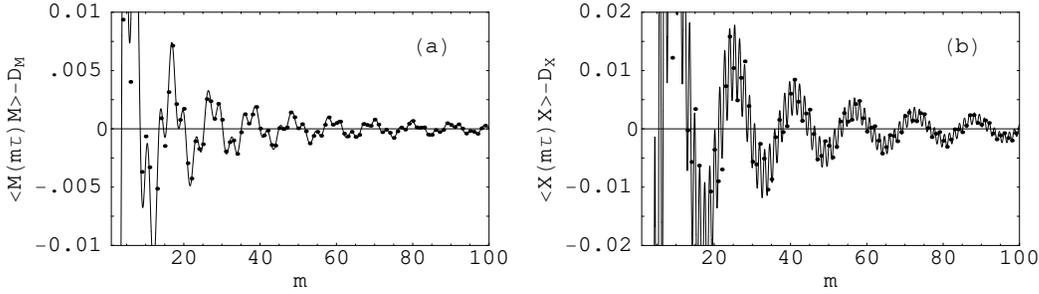}}
\vspace{-2.5mm}
\caption{The time-correlation functions (average subtracted) of (a) magnetization $M$ for $\alpha=0.35$,
$\beta=0.65$, and of (b) XX-energy $X$ for $\alpha=1.3$,$\beta=1.1$.
Dots are exact (numerical) results while thin solid curves are asymptotic formulae 
(\ref{eq:result},\ref{eq:magngen}).
}
\label{fig:tcor}
\end{figure}

Now we will establish that also spectral measures for the general case
(arbitrary kicking parameters $\alpha$,$\beta$) have square-root singularities
(see fig.\ref{fig:meas}) and thus lead to $t^{-3/2}$ decay of correlations.
We will first only consider contributions from the upper spectral band
$+\vartheta(\varphi)$, while the contribution of the lower
band is obtained simply by replacing $A$ by $A^\dagger$ (or $\ul{a}$ by $\ul{a}^*$).
Below we assume that $\alpha\ge 0$,$\beta\ge 0$,
so $|\alpha-\beta|\le|\alpha+\beta|$, while other cases can be obtained with 
trivial modifications.
Let us first expand the band around the minimum for small quasi-momentum
$\varphi=\varepsilon$, $\vartheta(\varepsilon)=2|\alpha-\beta| + 
\half (s_\alpha s_\beta/s_{|\alpha-\beta|})\varepsilon^2 + \Ord{\varepsilon^3}$.
Then the scattering data (\ref{eq:asansatz},\ref{eq:psi}-\ref{eq:psi0}) are
expanded explicitly in leading two orders 
\begin{eqnarray}
\vec{v}^\varepsilon &=& (-\ihalf,\ihalf,\sgn(\alpha\!-\!\beta)\shalf) -
\half \varepsilon \frac{e^{2i\sgn(\alpha\!-\!\beta)\alpha}-1}{e^{2i|\alpha-\beta|}-1} (1,1,0)
+\Ord{\varepsilon^2},\label{eq:scdat2}\\
S^\varepsilon &=& -1 + \Ord{\varepsilon^2}.\nonumber
\end{eqnarray}
When we evaluate the quasi-momentum densities (\ref{eq:rho}) we find that expression
inside $|.|$ vanishes to order $\Ord{1}$ so we have 
$\rho^+_A(\varepsilon) = K_A \varepsilon^2 + \Ord{\varepsilon^3}$
with coefficient
\begin{equation}
K_A := \frac{1}{2\pi}\left|
a_0 \frac{s_\alpha}{s_{\alpha-\beta}} +
\sum_{n=1}^\infty \vec{a}_n \cdot 
\left\{\frac{s_\alpha}{s_{\alpha-\beta}}(1,1,0) + n e^{i\sgn(\alpha\!-\!\beta)\beta} 
(-1,1,i\sgn(\alpha\!-\!\beta)\sqrt{2})\right\}
\right|^2
\end{equation}
Similarly we find, expanding around the other end (the maximum) of the band,
$\varphi=\pi-\varepsilon$, 
$\vartheta(\pi-\varepsilon)=
2|\alpha+\beta|-\half (s_\alpha s_\beta/s_{|\alpha+\beta|})\varepsilon^2
+\Ord{\varepsilon^2}$,
the scattering data
\begin{equation}
\vec{v}^\varepsilon = (\ihalf,-\ihalf,\shalf) -
\half \varepsilon \frac{e^{2i\alpha}-1}{e^{2i(\alpha+\beta)}-1} (1,1,0)
+\Ord{\varepsilon^2},\quad
S^\varepsilon = -1 + \Ord{\varepsilon^2},\quad
\label{eq:scdat3}
\end{equation}
and the quasi-momentum densities 
$\rho^+_A(\pi-\varepsilon) = L_A \varepsilon^2 + \Ord{\varepsilon^3}$
with 
\begin{equation}
L_A := \frac{1}{2\pi}\left|
a_0 \frac{s_\alpha}{s_{\alpha+\beta}} +
\sum_{n=1}^\infty (-1)^n \vec{a}_n \cdot 
\left\{\frac{s_\alpha}{s_{\alpha+\beta}}(1,1,0) + n e^{-i\beta} (-1,1,-i\sqrt{2})\right\}
\right|^2
\end{equation}
Transforming to spectral variable $\vartheta$ and multiplying by the density of states 
(\ref{eq:rho2mu}), which has a simple form 
$\propto |\vartheta-2|\alpha\pm\beta||^{1/2}$ around both respective ends of the 
spectral band, we obtain explicit square-root singularities of the spectral measure
\begin{eqnarray*}
\mu'_A(2|\alpha-\beta| + \xi) &=& 
\frac{1}{2}\left|\frac{s_{\alpha-\beta}}{s_\alpha s_\beta}\right|^{\frac{3}{2}}K_A 
\sqrt{\xi} + \Ord{\xi^{3/2}}, \cr
\mu'_A(2|\alpha+\beta| - \xi) &=& 
\frac{1}{2}\left|\frac{s_{\alpha+\beta}}{s_\alpha s_\beta}\right|^{\frac{3}{2}}L_A 
\sqrt{\xi} + \Ord{\xi^{3/2}},
\quad
\end{eqnarray*}
and similarly for the lower band by replacing $A$ by $A^\dagger$.
Since the spectral measure $\mu'_A(\vartheta)$ is a smooth-function on a complex 
unit-circle, except for four singularities at the four band edges $\pm 2\alpha \pm 2\beta$,
the asymptotic approximation to the integral (\ref{eq:tcor}) is dominated by the four
terms which are computed using asymptotics (\ref{eq:asymptotics}) (see fig.\ref{fig:tcor})
\begin{eqnarray}
\ave{A(m\tau)A^\dagger} \approx D_A + \frac{\sqrt{\pi}}{4}
\Bigl\{&&\left|\frac{s_{\alpha-\beta}}{s_\alpha s_\beta}\right|^{\frac{3}{2}}
\left(K_A e^{i(2|\alpha-\beta|m+\eta)} + K_{A^\dagger} e^{-i(2|\alpha-\beta|m+\eta)}\right) 
\nonumber\\
+&&\left|\frac{s_{\alpha+\beta}}{s_\alpha s_\beta}\right|^{\frac{3}{2}}
\left(L_A e^{i(2|\alpha+\beta|m-\eta)} + L_{A^\dagger}e^{-i(2|\alpha+\beta|m-\eta)}\right)
\Bigr\}|m|^{-\frac{3}{2}}\nonumber\\
{\rm with}\quad \eta:=(3\pi/4)\sgn\,m. \quad  && \label{eq:result}
\end{eqnarray}
In case of a hermitean operator $A=A^\dagger$ the formula
simplifies and then, of course, time-correlations are real and symmetric 
$\ave{A(m\tau)A}=\ave{A(-m\tau)A}=\ave{A(m\tau)A}^*$.
For example, the asymptotics for magnetization is (for $m>0$, again upto $\Ord{m^{-5/2}}$)
\begin{equation}
\ave{M(m\tau)M} \approx D_M + \frac{1}{4}\left|\frac{s_\alpha}{\pi s_\beta^3}\right|^{\frac{1}{2}}
\left\{\frac{\sin(2|\alpha\!+\!\beta|m\!-\!{\textstyle\frac{\pi}{4}})}{|s_{\alpha+\beta}|^{1/2}}
      -\frac{\sin(2|\alpha\!-\!\beta|m\!+\!{\textstyle\frac{\pi}{4}})}{|s_{\alpha-\beta}|^{1/2}}
\right\}m^{-\frac{3}{2}},
\label{eq:magngen}
\end{equation}
which {\em does not converge} to (\ref{eq:magn}) if we let $\tau=\alpha=\beta/J\rightarrow 0$ 
while keeping $t=m\tau$ large meaning very explicitly that the limits $\tau\rightarrow 0$ and $t\rightarrow\infty$ {\em do not commute}.

Note that all the quantities computed above are non-smooth functions of parameters 
$\alpha,\beta$ on the line $\alpha=\beta\pmod{\pi}$ (e.g, see fig.\ref{fig:stif}) 
since the band minimum $\vartheta_{\rm min}=2|\alpha-\beta|$ is non-smooth. 
For $\alpha=\beta$ the two bands touch (at the point $\vartheta=0$ where $\mu'_A(\vartheta)$
then becomes a smooth function), and the square-root singularities at 
$\pm 2|\alpha-\beta|=0$ dissapear, and so should also the two terms with 
$K_A,K_{A^\dagger}$ in (\ref{eq:result}).

\section{Conclusion}
The problem of infinite-temperature time-correlation functions in a family of 
Ising spin $1/2$ chains periodically kicked with transversal field, 
which has been formulated in terms of a spectral problem for the adjoint propagator 
over a certain subspace of observables in Heisenberg picture, has been solved 
using methods (and terminology) of a single-particle quantum scattering on a 
semi-infinite tight-binding lattice. It has been shown that time-autocorrelation 
function generally decays as $t^{-3/2}$ to 
its saturation value which is, due to integrability, 
generally different from the squared canonical average. 
This may be interpreted in terms of a relaxation to a 
{\em non-unique} equilibrium  statistical steady-state. Furthermore, 
it {\em excludes} the possibility of {\em (quasi)periodic motion}, 
which is a qualitatively {\em different} situation than the one we encounter in 
{\em few-body} classical (or quantum) integrable systems.
Note that such behaviour is drastically different from dynamics of some 
`trivially' integrable quantum many-body lattices, such as XX or 
Ising chain without external magnetic field (e.g., put $\beta=h=0$ in the results above) 
where the continuous spectrum of adjoint dynamics {\em collapses to a point} and one 
recovers {\em periodic} time-correlations.
It is an open challenge whether our approach can be extended to more 
`sophisticated' integrable quantum lattices, such as the general XYZ-chain or the 
Hubbard model, perhaps within a formalism of quantum inverse scattering.

I hope that these results may find some interesting application, and may help to 
stimulate some development of ergodic theory of quantum many-body systems.

\section*{Acknowledgements}
Hospitality of the Max Planck Institut f\" ur Kernphysik, Heidelberg, where this work has
been completed, and the financial support by the Ministry of science and technology of Republic 
of Slovenia are gratefully acknowledged.

\end{document}